\newif\iftwocolumnmode
\twocolumnmodefalse

\iftwocolumnmode
\documentclass[10pt]{iopart}
\else
\documentclass[12pt]{iopart}
\fi

\usepackage{graphicx}
\usepackage{cite}
\usepackage{siunitx}
\usepackage{color}
\usepackage{comment}
\usepackage{datetime}
\usepackage[%
  setpagesize=false,%
  bookmarks=true,%
  bookmarksdepth=tocdepth,%
  bookmarksnumbered=true,%
  colorlinks=true,%
  allcolors=magenta,%
  pdftitle={},%
  pdfsubject={},%
  pdfauthor={Kosuke Mizuno},%
  pdfkeywords={}%
]{hyperref}

\begin{document}

\title[]{Fast coherent control of nitrogen-14 spins associated with nitrogen-vacancy centers in diamonds using dynamical decoupling}

\author{Kosuke Mizuno$^{1 \dagger}$, Ikuya Fujisaki$^1$, Hiroyoshi Tomioka$^1$, Hitoshi Ishiwata$^{1,2 \ddagger}$, Shinobu Onoda$^3$, Takayuki Iwasaki$^1$, Keigo Arai$^{1,2}$, and Mutsuko Hatano$^{1 \ast}$}
\address{$^1$ School of Engineering, Tokyo Institute of Technology, Tokyo 152-8550, Japan}
\address{$^2$ Japan Science and Technology PRESTO, Saitama 332-0012, Japan}
\address{$^3$ National Institutes for Quantum Science and Technology, 1233 Watanuki, Takasaki, Gunma 370-1292, Japan}
\address{$^\dagger$ Present address: NTT Basic Research Laboratories, Kanagawa 243-0198, Japan}
\address{$^\ddagger$ Present address: National Institutes for Quantum and Radiological Science and Technology, Chiba 263-8555, Japan}
\ead{$^\ast$hatano.m.ab@m.titech.ac.jp}

\begin{abstract}
  A nitrogen-vacancy (NV) center in a diamond enables the access to an electron spin, which is expected to present highly sensitive quantum sensors.
  Although exploiting a nitrogen nuclear spin improves the sensitivity, manipulating it using a resonant pulse requires a long gate time owing to its small gyromagnetic ratio.
  Another technique to control nuclear spins is a conditional rotation gate based on dynamical decoupling, which is faster but unavailable for nitrogen spins owing to the lack of transverse hyperfine coupling with the electron spin.
  In this study, we generated effective transverse coupling by applying a weak off-axis magnetic field.
  An effective coupling depends on the off-axis field; the conditional rotation gate on the nitrogen-14 spins of an NV center was demonstrated within \SI{4.2}{\micro\second} under an \SI{1.8}{\percent} off-axis field and a longitudinal field of approximately \SI{280}{\milli\tesla}.
  We estimated that a population transfer from the electron to nitrogen spins can be implemented with \SI{8.7}{\micro\second}.
  Our method is applicable to an ensemble of NV centers, in addition to a single NV center.
\end{abstract}

\vspace{2pc}

\iftwocolumnmode
  \ioptwocol
\else
\fi

\section{Introduction}

A nitrogen-vacancy (NV) center in a diamond is a point defect with an isolated electron spin, which is utilized as a highly sensitive quantum magnetometer under ambient conditions~\cite{Taylor2008, Wolf2015PRX, CZhang2021PRAppl}.
It has received considerable attention for biological applications~\cite{LeSage2013, Barry2016, McCoey2020SM, Igarashi2020JACS, Fujiwara2021NT, Ishiwata2022AQT, Arai2022CP}, biomedical applications~\cite{Kuwahata2020SR}, applied physics~\cite{YHatano2021APL, Sturner2019}, and material physics\cite{Zhou2019PRX, Ku2020graphene}.
NV centers are appealing owing to their applicability; however, they are less sensitive than optically pumped magnetometers and superconducting interference devices~\cite{RevModPhys.92.021001}.
The limiting factor for the sensitivity of NV-based sensors is the coherence time of the electron spin, which is generally a few tens of microseconds.
Storing the information of the sensor spins in a long-lasting memory is one of the solutions for improving sensor characteristics; for example, the nitrogen atom of an NV center has a spin freedom of degree with a significantly longer coherence time.
Utilizing the nitrogen spin as the quantum memory has been demonstrated to extend the capabilities of individual NV centers~\cite{Rosskopf2016, Lovchinsky2016, Zaiser2016, Pfender2017, Aslam2017}; in particular, Lovchinsky~\textit{et al.}~\cite{Lovchinsky2016} achieved an order improvement in sensitivity by repetitive readout.

These nuclear-assisted protocols require nuclear spin manipulations.
Although nuclear spins can be simply controlled by irradiating a resonant radio-frequency (RF) pulse, it requires long gate times, typically a few hundred microseconds~\cite{Zaiser2016}, owing to their gyromagnetic ratios being as small as approximately one-thousandth of the electron spin.
An alternative method is an electron-conditional nuclear spin-rotation gate~\cite{Taminiau2014} designed with dynamical decoupling~\cite{de2010universal}.
Dynamical decoupling consisting of multiple $\pi$ pulses on the electron spin modifies magnetic noises that the electron undergoes~\cite{Kotler2011, de2010universal, Souza2011, Bar-Gill2012NC, Bar-Gill2013NC}, enabling the frequency-selective detection of magnetic field~\cite{Hall2010aPRB, deLange2011PRL, Pham2012PRB} and selective enhancement of coupling to nuclear spins~\cite{Staudacher2013, Muller2014, Rugar2014, Haberle2015, Lovchinsky2017, Ishiwata2017APL, Sasaki2020aAPL}.
It has also been demonstrated for controlling proximal carbon-13 spins within a few microseconds~\cite{Taminiau2012, Kolkowitz2012, Taminiau2014, GQLiu2017, Abobeih2018}.
This method utilizes the electron-nuclear spin coupling to control nuclear spins instead of direct RF irradiation, leading to simplicity in an experimental setup.
Although the conditional rotation gate relies on transverse coupling between the electron and target spins, the nitrogen spin does not exhibit this type of coupling with electrons.

Liu~\textit{et al.}~\cite{YXLiu2019} presented the perturbation theory of a weak off-axis magnetic field involving effective transverse coupling and demonstrated an off-axis field sensing.
The sensitivity to off-axis fields could be large by the ground state level anti-crossing (GS-LAC).
In this study, we generated effective transverse coupling by applying an off-axis field to demonstrate that the conditional rotation gate on the nitrogen spin within a few microseconds.
Moderate effective coupling strength is still available far from the GSLAC, enabling fast coherent operations of the nuclear spin. 
We applied a magnetic field of \SI{280}{\milli\tesla} with an off-axis field of \SI{5}{\milli\tesla} and observed coherent oscillations of the nitrogen-14 spin.
The effective coupling strength was tunable within 10--\SI{90}{\kilo\hertz} by varying the off-axis field within 2--\SI{7}{\milli\tesla}.
Using our method, the estimated gate time for a population transfer from the electron to the nuclear spin was \SI{8.7}{\micro\second}.
We conducted the same experiments on an ensemble of NV centers, for which the proposed method rapidly controlled the nitrogen spins.
The proposed method enables the formation of nuclear-assisted protocols for an ensemble of NV centers.
Since this method is free from RF pulses and thus requires no additional instruments like proximal RF antennas, it could be a way to integrate nuclear-assisted protocols into wide-field~\cite{Fu2014Sci, Fescenko2018PRAppl, Lillie2019PRAppl, Mizuno2020, McCoey2020SM, Turner2020PRAppl, Kitawaga2023PRAppl} or large ensemble systems~\cite{Barry2016, Schloss2018PRAppl, Hatano2022SR, KUBOTA2023109853} that are important for practical sensing applications.

The rest of the paper is organized as follows.
Section~\ref{sec:principle} describes NV centers and the principle of our method.
Section~\ref{sec:method} presents details of an experimental setup and samples.
Section~\ref{sec:result} shows experimental results.
Section~\ref{sec:discussion_comparison} provides a comparison of our method to previous studies.
Section~\ref{sec:discussion_requirements} describes experimental requirements of the method.
Finally, Section~\ref{sec:discussion_transfer} discuss an application for nuclear-assisted protocols.

\section{Principle of the nitrogen spin manipulation} \label{sec:principle}

\begin{figure*}[htbp]
  \centering
  \includegraphics[width=.75\textwidth]{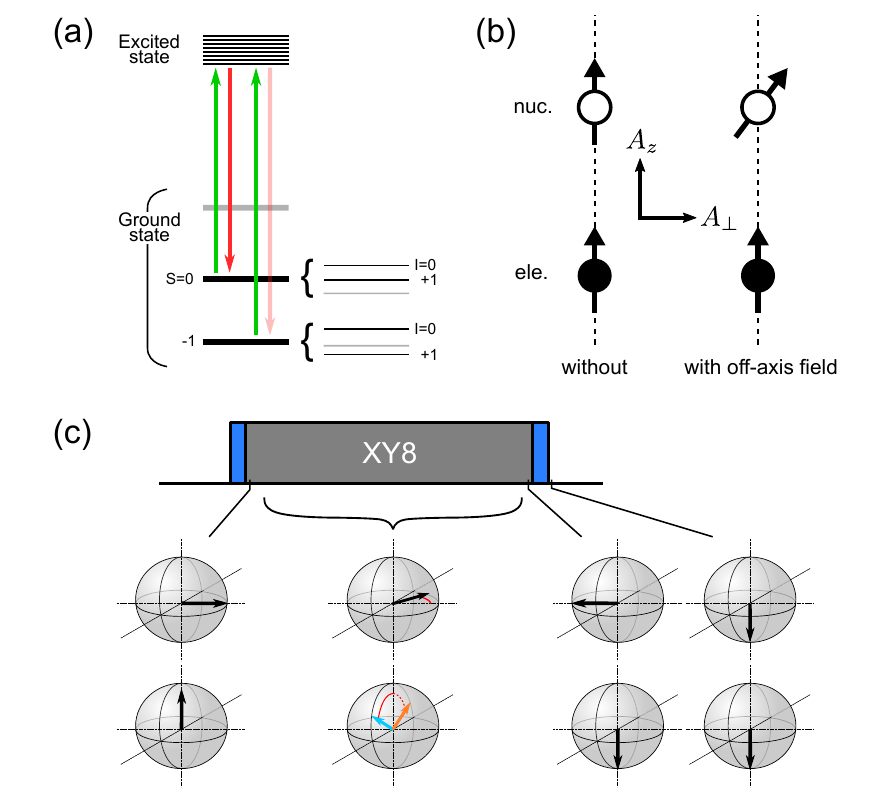}
  \caption{
    (a)
    An energy diagram of a $^{14}\mathrm{N}$--V center in a diamond.
    The subspace spanned by $S=\lbrace -1, 0 \rbrace$, and $I=\lbrace 0, +1 \rbrace$ are emphasized for the electron and nitrogen-14 spins, respectively.
    (b)
    The hyperfine interaction between the electron and nuclear spins does not include a transverse coupling $\hat S_z \hat I_\perp$, which is required for electron-conditional nuclear spin rotation gates based on dynamical decoupling.
    A weak off-axis magnetic field perturbatively generates an effective transverse coupling.
    (c)
    An example of a nuclear spin control.
    The effective transverse interaction $g \hat S_z \hat I_\perp$ under dynamical decoupling rotates the electron and nuclear spins in a conditioned fashion, which allows the universal control of the electron-nuclear spin system.
  }
  \label{fig:principle}
\end{figure*}

Fig.~\ref{fig:principle}(a) presents the structure of an NV center in a diamond and its energy diagram.
An NV center consists of a substitutional nitrogen atom and an adjacent vacancy that localizes an electron $S=1$ system.
The spin states of the NV centers are initialized and detected using laser illumination.
The Hamiltonian of this electron system is as follows:
\begin{equation}
  \hat H_0 = D_{gs} \hat S_z + \gamma_e (B_z \hat S_z + B_\perp \hat S_\perp),  \label{eq:H0}
\end{equation}
where we focus on a subspace of $S=\lbrace-1, 0\rbrace$~\cite{YXLiu2019}, $D_{gs}/2\pi=\SI{2.87}{\giga\hertz}$ is the zero field splitting of the NV centers, and $\gamma_e/2\pi=\SI{28}{\giga\hertz/\tesla}$ is the gyromagnetic ratio of the electron.
$\hat S_z$ and $\hat S_\perp$ correspond to the spin operators along the NV center axis ($Z$ axis) and the spin operator in the perpendicular direction, respectively.
A magnetic field can be decomposed into an axial component $B_z$ and an off-axis component $B_\perp$.
An electron undergoes hyperfine coupling to the nitrogen nuclear spin, and the total Hamiltonian is as follows:
\begin{eqnarray}
  \hat H &= \hat H_0 + \omega_n \hat I_z + \hat H_\mathrm{hf}, \quad\mathrm{with} \\
  \hat H_\mathrm{hf} &= A_z \hat S_z \hat I_z + A_\perp \left( \hat S_x \hat I_x + \hat S_y \hat I_y \right),  \label{eq:H_hf}
\end{eqnarray}
where $\omega_n$ is the transition energy of the nuclear spin, $\hat I_i\,(i=x,y,z)$ is the nuclear spin operator, and $A_z$ and $A_\perp$ correspond to the hyperfine coupling strengths of the longitudinal and transverse axes, respectively.
The longitudinal hyperfine interaction (the first term in Eq.(\ref{eq:H_hf})) is canceled by dynamical decoupling, and the transversal hyperfine interaction (the second term in Eq.(\ref{eq:H_hf})) is suppressed by the rotating wave approximation.
Although the nitrogen spin thus has no effect on the electron spin, a weak off-axis field $B_\perp$ rotates the quantization axis and generates an effective transverse coupling $\hat S_z \hat I_\perp$, which can be utilized with dynamical decoupling (Fig.~\ref{fig:principle}(b)).
Thus, the hyperfine interaction becomes the following:
\begin{equation}
  \hat H_\mathrm{hf}^\prime = A_z \hat S_z \hat I_z + \frac{\gamma_e B_\perp A_\perp}{D_{gs} - \gamma_e B_z} F \, \hat S_z \hat I_\perp ,  \label{eq:H_hf_eff}
\end{equation}
where $F$ is a constant determined from the second-order perturbation theory~\cite{YXLiu2019}.

The second term in Eq.(\ref{eq:H_hf_eff}) oscillates with the frequencies $\omega_n$ in a rotating frame with $\omega_n \hat I_z$.
Since dynamical decoupling techniques can modify the magnetic spectra the electron spin undergoes, such an oscillating term can be turned on and off.
The magnetic sensitivity under dynamical decoupling is maximized at a frequency $f_\mathrm{DD}$ such that $1/2 f_\mathrm{DD} = 2\tau$, where $2\tau$ is the spacing of $\pi$ pulses.
Thus, the NV center undergoes the second term under dynamical decoupling with a resonant condition $(1/2) \cdot (\omega_n \pm A_z/2) / 2\pi = 2\tau$; otherwise, it is decoupled from the second term.
The deviations $\pm A_z / 2$ from the resonant condition come from the longitudinal hyperfine coupling.

Note, there are two isotopes of nitrogen: $^{14}\mathrm{N}$ ($I=1$) and $^{15}\mathrm{N}$ ($I=1/2$).
We used $^{14}\mathrm{N}$ because it has a quadrupole interaction, resulting in a higher $\omega_n$ and thus frequent $\pi$ pulses in dynamical decoupling, which achieves better coherence protection.
For nitrogen-14, $A_z/2\pi=\SI{2.2}{\mega\hertz}$, $A_\perp/2\pi=\SI{-2.62}{\mega\hertz}$, and $F \simeq 2.75$~\cite{YXLiu2019}.
Moreover, we focus on the subspace of $I=\lbrace 0, +1 \rbrace$.
Note that the nuclear spin was not actively initialized because it was not required for our experiments. 

Instantaneous dynamics under dynamical decoupling are complicated.
Thus, we focus on the time-averaged Hamiltonian in the rotating frame as follows:
\begin{eqnarray}
  \bar H &= g \hat S_z \hat I_\perp, \quad \mathrm{with} \label{eq:H_av} \\
  g &= \frac{\gamma_e B_\perp A_\perp}{\pi (D_{gs} - \gamma_e B_z)} F,  \label{eq:g_eff}
\end{eqnarray}
where $g$ is the effective coupling strength.
Fig.~\ref{fig:principle}(c) demonstrates an example of the time-averaged dynamics of an NV center undergoing dynamical decoupling with the resonant condition.
Starting from an initial state $|S=0, I=+1\rangle$, the first $\pi/2$ pulse makes a superposition state of the electron spin.
The electron superposition state undergoes the Hamiltonian (Eq.(\ref{eq:H_av})) and rotates around the $Z$-axis of the Bloch sphere.
The nuclear spin simultaneously rotates around the $\pm X$axes in a conditioned fashion with respect to the electron spin.
Denoting the number of $\pi$ pulses of the dynamical decoupling as $N_p$ and the interaction time as $T = 2 \tau N_p$, the rotation angle $\theta = g T$ can be tuned by the following three parameters: $B_z$, $B_\perp$ and $N_p$.
When $\theta = \pi$, the state after the second $\pi/2$ pulse is $|S=0, I=0\rangle$; thus, this sequence is an $X$ gate of the nuclear spin.

\begin{figure*}[htbp]
  \centering
  \includegraphics[width=.9\textwidth]{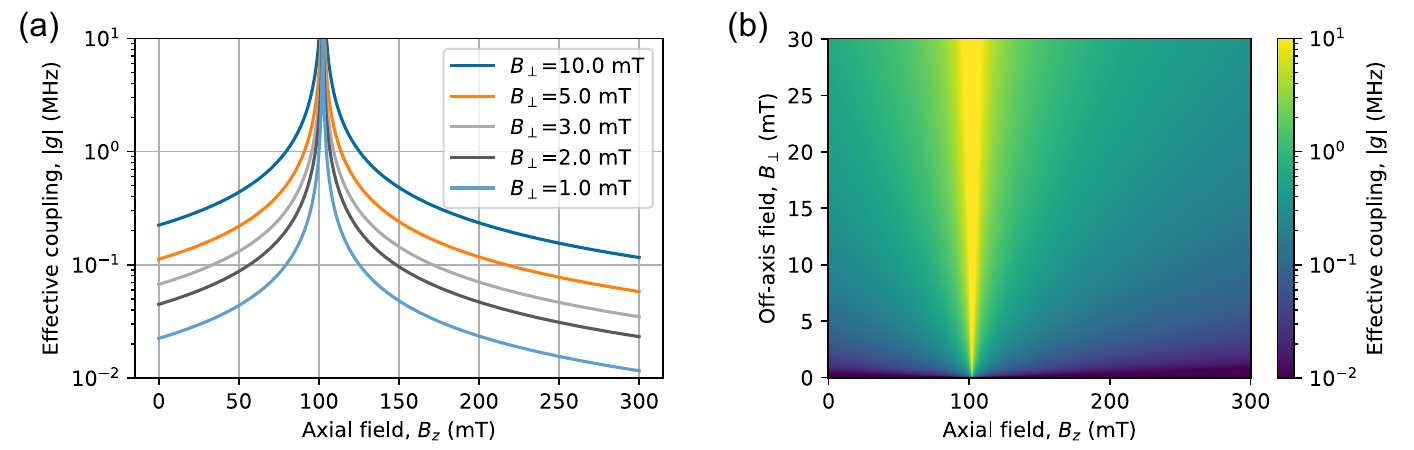}
  \caption{
    Effective coupling strength
    (a) plotted as a function of the axial field $B_z$ for several off-axis fields $B_\perp$.
    (b) Two dimentional plot for $(B_z, B_\perp)$.
    While the effective coupling strength diverges around the GSLAC, a moderate coupling strength is available for a strong field regime far from the GSLAC.
  }
  \label{fig:coupling}
\end{figure*}

Fig.~\ref{fig:coupling} shows an effective coupling strength calculated by Eq.(\ref{eq:g_eff}).
While a large coupling is available around the GSLAC at \SI{102.4}{\milli\tesla} due to the divergence, a moderate coupling strength is still available far from the GSLAC.
Assuming $B_z = \SI{250}{\milli\tesla}$ and $B_\perp=\SI{1}{\milli\tesla}$, the coupling strength is larger than \SI{10}{\kilo\hertz}, enabling a fast conditional $\pi$ rotation within \SI{5}{\micro\second}.

\section{Materials and Method} \label{sec:method}

\begin{figure*}[htbp]
  \centering
  \includegraphics[width=.90\textwidth]{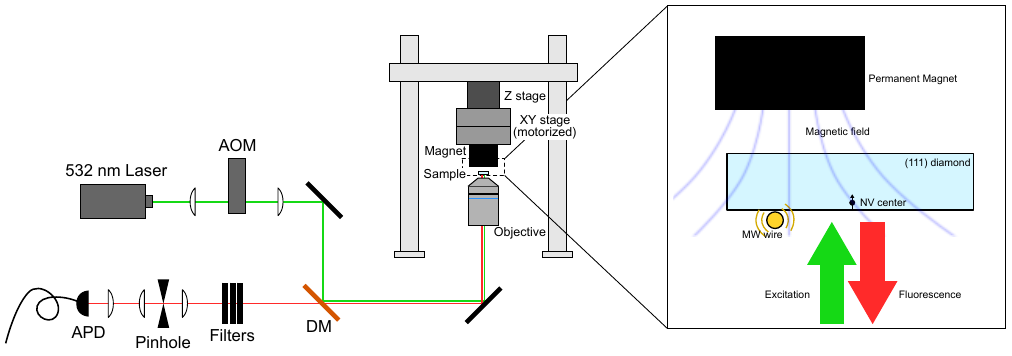}
  \caption{
    A home-built confocal microscopy.
    (111) diamond substrates were used.
    A suspended permanent magnet applied a magnetic field along with the NV center axes.
    A transverse magnetic field was controlled by changing the relative planar position between the magnet and NV centers at the focal spot.
  }
  \label{fig:setup}
\end{figure*}

A confocal microscope was used in this study (Fig.~\ref{fig:setup}).
Laser pulses at \SI{532}{\nano\metre} were chopped by an acousto-optic modulator (AOM) and irradiated to NV centers through an objective lens.
The fluorescence from the NV centers was collected by the objective, selected through optical filters and a pinhole, and detected by avalanche photodiodes (APDs).
We suspended a permanent magnet with vertical magnetization by applying a strong magnetic field around \SI{280}{\milli\tesla} along the $Z$-axis of the NV centers on the (111) diamond substrates.
A permanent magnet was installed with stepping motors.
We controlled the off-axis field by displacing the planar position of the magnet.
Microwave pulses for the electron spin control were generated using digital IQ modulation with a data timing generator (Tektronix, DTG5274), amplified, and then irradiated via a copper wire.
The typical length of a $\pi$ pulse on an electron spin is \SI{35}{\nano\second}.
Note that the length of the $\pi$ pulses was ignored in the previous section for simplicity.
For a finite $\pi$ pulse length, $t_\pi > 0$, the resonant condition must be modified to $1/2f_\mathrm{DD} = 2\tau + t_\pi$.
Several types of dynamical decoupling are theoretically equivalent.
We used the XY8~\cite{Gullion1990} sequence owing to its simple implementation and robustness against experimental imperfections.

Although the fluorescence intensity of the NV centers reflects the electron spin population, several unwanted signals (background light, shot noise in the detectors, and charge dynamics of the NV centers) were also included.
To extract the electron spin population from the fluorescence intensity, the fluorescence intensities were normalized in this study.
We conducted a measurement sequence, such as that shown in Fig.~\ref{fig:principle}(c) and recorded the raw fluorescence intensity $S_0$.
We also conducted an additional measurement by substituting the phase of the last $\pi/2$ pulse into the opposite phase ($-X/2$ pulse was used instead of $X/2$) and recorded the raw fluorescence intensity $S_1$.
The (normalized) fluorescence intensity was then calculated as $S = (S_0 - S_1)/(S_0 + S_1)$, mitigating the effects of the unwanted signals.

Two diamond substrates were prepared.
A IIa(111) diamond substrate containing individual NV centers was fabricated by implanting $^{15}\mathrm{N}^{+}$ ions accelerated at \SI{40}{\kilo\electronvolt} with a dose of \SI{5e8}{{cm}^{-2}} and vacuum annealing at \SI{1000}{\celsius} for \SI{2}{\hour}.
Although the implanted ion was $^{15}\mathrm{N}^{+}$, this sample contained several $^{14}\mathrm{N}$--V centers.
The carbon atoms in the substrate were naturally abundant; thus, the NV centers potentially had several carbon-13 spins in their vicinity.
An individual NV center with no strongly coupled carbon-13 spin ($< \SI{1}{\mega\hertz}$) was used, which was verified the optically detected magnetic resonance and the Ramsey experiments.
On the other hand, a IIa(111) diamond substrate containing high-density perfectly aligned NV centers was fabricated by chemical vapor deposition.
Although we did not investigate the nitrogen distribution in this sample, the growth conditions were nearly identical to those in a previous study\cite{Ishiwata2017APL}.
The nitrogen concentration should be \SI{170}{ppm} and confined in a shallow layer within \SI{10}{\nano\metre} from the surface.
Approximately one hundred NV centers were included in the detection spot of the confocal microscope.
The sample was grown using an isotopically purified carbon source ($[^{12}\mathrm{C}] \le \SI{99.995}{\percent}$).

\section{Results} \label{sec:result}

\subsection{Experiments on a single NV center} \label{sec:result_single}

\begin{figure*}[htbp]
  \centering
  \includegraphics[width=.90\textwidth]{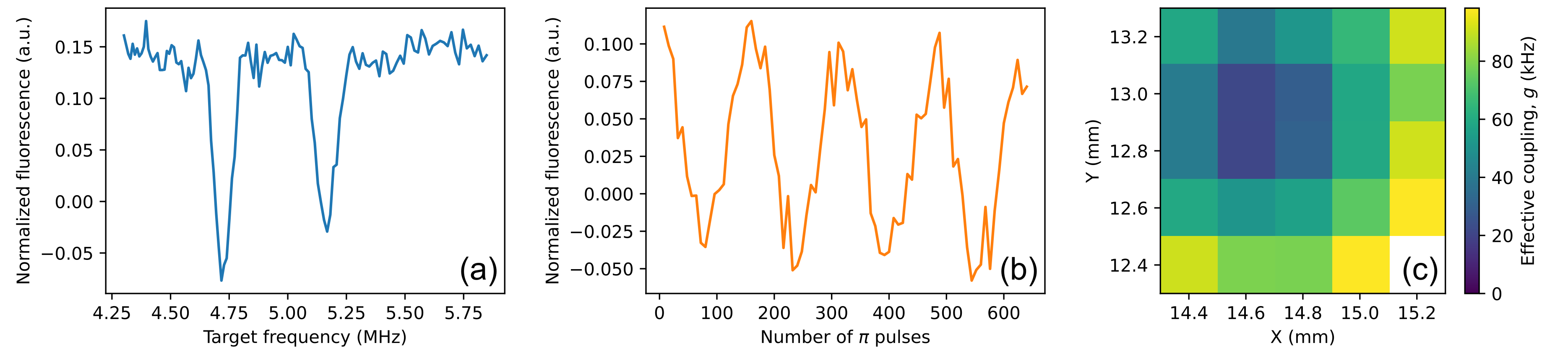}
  \caption{
    XY8 measurement of a single NV center.
    (a)
    Target frequency of the decoupling sequence versus fluorescence intensity with 80 pulses.
    (b)
    Number of pulses versus fluorescence intensity at the resonant condition of the left peak in (a), indicating a coherent oscillation of the nitrogen nuclear spin.
    (c)
    Effective coupling strength at several planar positions of the magnet.
  }
  \label{fig:result_single}
\end{figure*}

We demonstrated the generation of effective transverse coupling to the nitrogen spin of a single NV center.
As indicated in the Principle Section (Fig.~\ref{fig:principle}(c)), the superposition state of the electron spin after the first $\pi/2$ pulse rotates under an XY8 sequence if the resonant condition is satisfied, and the second $\pi/2$ pulse maps the rotation angle to the electron spin population.
Thus, with a small $N_p$, we can verify the existence of transverse coupling at a pulse spacing $\tau$ and at the corresponding resonant condition $f_\mathrm{DD}$.
Fig.~\ref{fig:result_single}(a) demonstrates the fluorescence intensity of the XY8 measurement of a single NV center for several $f_\mathrm{DD}$.
The number of $\pi$ pulses were $N_p=80$.
Two peaks were observed; the left (right) peak corresponds to the transition between $I=\lbrace +1, 0 \rbrace$ ($I=\lbrace 0, -1 \rbrace$).
Note, quantum interpolation~\cite{Ajoy2017} was used for Fig.~\ref{fig:result_single}(a) to obtain a fine frequency resolution.
To determine the effective transverse coupling strength, we measured the fluorescence intensity with several $N_p$ values under the resonant condition of the left peak (Fig.~\ref{fig:result_single}(b)).
Coherent oscillations of the nuclear spin were obtained.
The oscillation period was $N_p \simeq 160$, corresponding to an effective coupling strength of $g \simeq \SI{59}{\kilo\hertz}$, and the off-axis field was approximately \SI{5}{\milli\tesla}.
For the tuning of an effective coupling, we conducted these experiments with different positions of the permanent magnet.
Fig.~\ref{fig:result_single}(c) demonstrates the effective coupling strength as a function of the planar position of the permanent magnet.
Effective couplings of 10--\SI{90}{\kilo\hertz} were observed, corresponding to off-axis fields of 2--\SI{7}{\milli\tesla}.
The map of the effective coupling appears as a bowl, and the bottom of the bowl at $(X,Y)=(14.6, 13.0)$ may correspond to a point where the NV center axis coincides with the field direction of the permanent magnet.
Therefore, we used an off-axis magnetic field to generate an effective transverse hyperfine coupling.

\begin{figure*}[htbp]
  \centering
  \includegraphics[width=.90\textwidth]{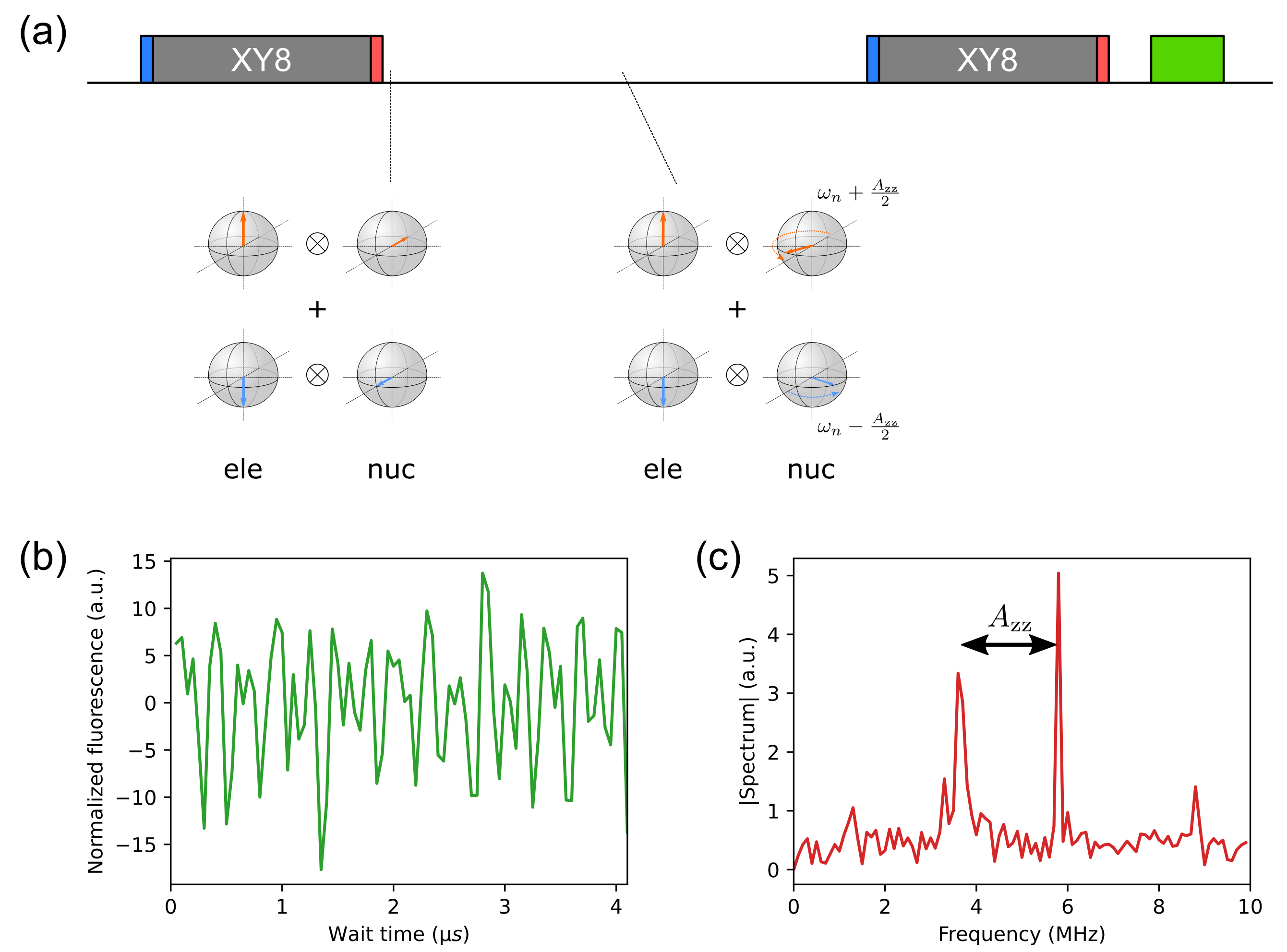}
  \caption{
    (a)
    XY8 correlation measurement of the single NV center.
    Each XY8 was conducted with a resonant condition; the rotation angle was $\pi/2$.
    The first XY8 entangles the electron and nuclear spins.
    The nuclear spin rotates with electron-dependent Larmor frequencies $\omega_n \pm A_z / 2$ during the free evolution time, which is owing to the longitudinal hyperfine coupling.
    The second XY8 maps the rotation angle back to the electron population.
    (b,c)
    Fluorescence intensity versus the evolution time (b) and its spectrum (c).
    The two peaks, which exhibited a separation of \SI{2.2}{\mega\hertz}, coincided with the longitudinal hyperfine coupling.
  }
  \label{fig:result_single_corr}
\end{figure*}

A correlation measurement was conducted to confirm that effective coupling can be utilized for the conditional rotation gate.
Fig.~\ref{fig:result_single_corr}(a) demonstrates the pulse sequence of the correlation measurement composed of an XY8 sequence, free evolution time, and XY8 sequence.
The $X/2$ pulse (blue) prepares the electrons in a superposition state.
The XY8 sequence with a resonant condition for the nuclear spin (gray), for which the rotation angle is $\pi/2$, entangles the electron and nuclear spins.
The $Y/2$ pulse (red) turns the electron spin back into the $Z$-axis of the Bloch sphere.
The resultant state before the free-evolution time is an entangled state of the electron and nuclear spins.
During the free evolution time, the nitrogen spin rotates with two Larmor frequencies owing to the longitudinal hyperfine coupling between the electron and nitrogen spins.
The frequency difference corresponds to the longitudinal coupling $A_z/2\pi$ (the first term in Eq.(\ref{eq:H_hf_eff})).
The second XY8 sequence between the $X/2$ and $Y/2$ pulses translates the rotation angle of the nuclear spin back into the electron spin population~\cite{Staudacher2015}.
Fig.~\ref{fig:result_single_corr}(b) demonstrates the fluorescence intensity of this measurement; its frequency spectrum is shown in Fig.~\ref{fig:result_single_corr}(c).
The number of XY8 sequences was $N_p = 40$, and the expected rotation angle was approximately $\pi/2$.
Two peaks with a separation of \SI{2.2}{\mega\hertz} were observed, which coincides with the longitudinal hyperfine coupling $A_z/2\pi$.
This result indicates that XY8 sequences with a resonant condition rotate the nitrogen spin, achieving an expected conditional rotation gate.

\subsection{Experiments on ensemble NV centers} \label{sec:result_ensemble}

\begin{figure}[htbp]
  \centering
  \includegraphics[width=.90\textwidth]{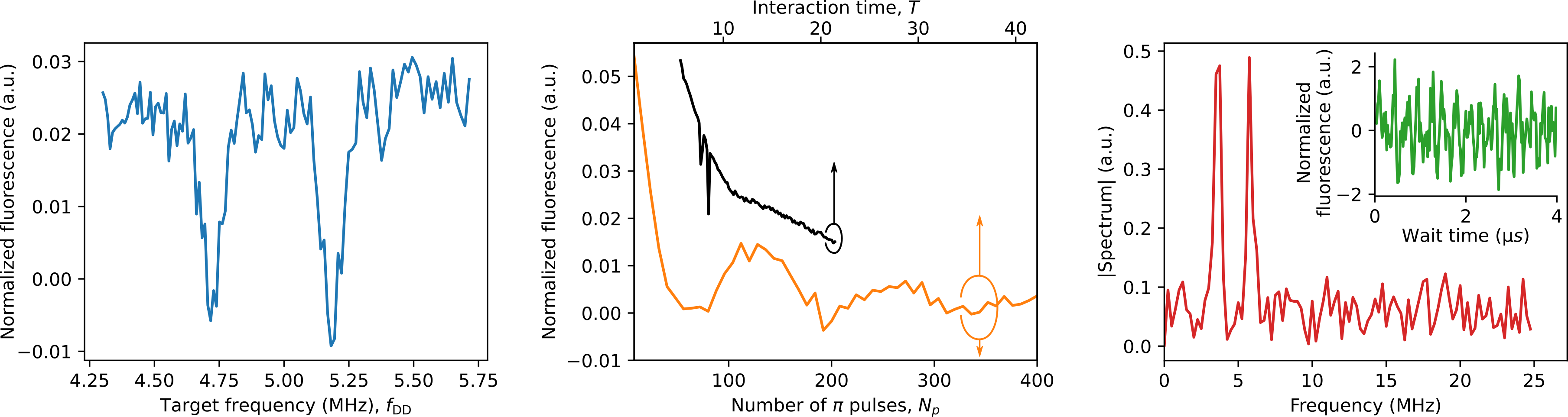}
  \caption{
    Measurements with an ensemble of NV centers.
    (a)
    An XY8 measurement with varying target frequencies.
    (b)
    Orange: A coherent oscillation at the left peak of (a). Black: an XY8 measurement with varying pulse spacings with 80 pulses.
    (c)
    a spectrum of the correlation measurement exhibiting a separation of \SI{2.2}{\mega\hertz}.
    Inset: time domain data.
  }
  \label{fig:result_ensemble}
\end{figure}

The same experiments were conducted on the ensemble of NV centers.
Fig.~\ref{fig:result_ensemble}(a) demonstrates the fluorescence intensity of an XY8 measurement as a function of $f_\mathrm{DD}$
The number of $\pi$ pulses was $N_p = 80$.
Fig.~\ref{fig:result_ensemble}(b, orange) demonstrates the fluorescence intensity of an XY8 measurement as a function of $N_p$ at the resonant condition to the left peak shown in Fig.~\ref{fig:result_ensemble}(a), indicating a coherent oscillation.
The oscillation of the ensemble of NV centers decayed more quickly than that of the single NV center, which may be attributed to the difference in the coherence times.
The black curve shown in Fig.~\ref{fig:result_ensemble}(b) presents the fluorescence intensity of an XY8 measurement with $N_p = 80$ pulses, indicating a coherence time of approximately \SI{10}{\micro\second}.
Fig.~\ref{fig:result_ensemble}(c) demonstrates the frequency spectrum of the correlation measurements of the ensemble of NV centers.
Two peaks with a separation of \SI{2.2}{ \mega\hertz } were also observed.
This indicates that our method, which generates an effective coupling by an off-axis field and controls the nuclear spin by dynamical decoupling, can be applied to an ensemble of NV centers.

\section{Discussions} \label{sec:discussion}

\subsection{Comparison to related studies} \label{sec:discussion_comparison}

Our work will be compared to previous studies.

Nuclear spins can be controlled directly by resonant RF pulses, however, requiring strong pulses~\cite{Arunkumar2022} or tailored RF antennas~\cite{Herb2020RSI}.
Otherwise, long gate operations are required, being detrimental due to the short coherence time of electron spins.
Some previous works have developed a combined strategy that simultaneously decouples an electron spin from the environment and manipulates a nuclear spin by an RF pulse and dynamical decoupling sequence~\cite{VanderSar2012, PhysRevX.9.031045, Abobeih2019}.

Meanwhile, our method needs no RF drive for nuclear spins.
Although it requires precise control of magnetic fields, as discussed below, it is free from proximal structures like lithographic patterns or RF antennas, preventing the applicability of NV centers.

Besides the above straightforward ways, hyperfine interaction between an electron and a nuclear spin could enhance the effective gyromagnetic ratio of the nuclear spin, resulting in fast manipulations of nuclear spins~\cite{PhysRevA.80.050302, PhysRevB.92.020101, PhysRevB.105.L020405, Degen2020, PhysRevLett.124.153203}.
This effect is maximized at the GSLAC condition for NV centers.
Note that a larger gyromagnetic ratio simultaneously means faster dephasing~\cite{Sangtawesin_2016}.
Our experiments are demonstrated in a strong field region ($B_z \sim \SI{280}{\milli\tesla}$, corresponding to $|D_{gs} - \gamma_e B_z| \simeq 2\pi \times \SI{5}{\giga\hertz}$) far from the GSLAC.
This allows us to avoid GSLAC-induced nuclear spin dephasing.
Moreover, our method is available in broad magnetic field conditions except for the GSLAC because Eq.(\ref{eq:g_eff}) assumes that $D_{gs} - \gamma_e B_z$ is large~\cite{YXLiu2019}.
This applicability is advantageous for practical sensing applications.

Although our work is inspired by Liu~\textit{et al.}~\cite{YXLiu2019}, which utilized the GSLAC ($|D_{gs} - \gamma_e B_z| \simeq 2\pi \times \SI{100}{\mega\hertz}$) to increase sensitivity, we demonstrate that the off-axis field can be utilized to control nuclear spin even in such a strong field condition.

\subsection{Experimental requirements}  \label{sec:discussion_requirements}

There are several limitations and experimental requirements for the conditional rotation gates on the nitrogen spins.
First, generating a desired transverse effective coupling is crucial for our method to achieve a conditional rotation gate with a desired rotation angle.
The rotation angle can be tuned by the number of pulses and effective coupling strength; however, the tunability is discretized based on the number of pulses.
Thus, precise control and stabilization of the magnetic field are particularly important.
Second, the timing resolution of the microwave pulses, which was \SI{0.5}{\nano\second} in our setup, confines the pulse spacing to discrete values, limiting the tunability of the rotation angle.
Quantum interpolation~\cite{Ajoy2017} or shaping the pulse waveform~\cite{Zopes2017} will help tune the rotation angle with a higher resolution.
Optimal control~\cite{Khaneja2005, Dolde2014, Tabuchi2017PRA} can be another solution for achieving nuclear spin manipulations via effective transverse coupling.

Note that the rotation angle is proportional to the effective coupling strength and thus proportional to the off-axis field $B_\perp$ (see Eq.(\ref{eq:g_eff})).
This means that a small field deviation s.t. $g\rightarrow g (1 + \epsilon)$ causes a quadratic error in fidelity~\cite{WIMPERIS1994221}.
Thus carefully designing an experimental setup could mitigate experimental imperfections, for example, deviation, temporal instability, spatial inhomogeneity, and gradient of the magnetic field.
Additionally, composite pulses~\cite{WIMPERIS1994221, Aiello2013NC, PhysRevLett.112.050503} and robust quantum control~\cite{PhysRevLett.112.050502, Ziem2019, Ball2021, Carvalho2021} could help further mitigation for such imperfections.

Our method can be applied to various experiments, including wide-field and large detection volume setups that are significant for actual sensing applications~\cite{Aslam2023}, owing to the simplicity of the electrical components.
Such setups are based on an ensemble of NV centers which exhibits much shorter  dephasing time than a single NV center.
Although the dephasing time of the electron spins should limit the fidelity of the conditional rotation gates, this time corresponds to a dephasing time $T_\mathrm{2,DD}$ under dynamical decoupling sequences rather than an inhomogeneous dephasing time $T_2^\ast$.
Since $T_\mathrm{2,DD}$ can be significantly longer than $T_2^\ast$ for high-density samples, our method is expected to be applicable to high-density NV centers.

\subsection{Nuclear-assisted sensing}  \label{sec:discussion_transfer}

Finally, the conditional rotation gate helps implement nuclear-assisted protocols; for example, a repetitive readout with the nitrogen spin as a quantum memory~\cite{Lovchinsky2016}.
Such nuclear-assisted protocols need a population transfer between the spins.
The conditional rotation gates enable to implementation of such operations.
Note that the conditional rotation gate is necessary rather than conditional phase gates~\cite{Sangtawesin2014PRL} to implement population transfer.
The latter gates additionally require direct control of nuclear spins.
Implementations of this operation include a conditional $\pi/2$ rotation gate, an unconditional $Z$ gate, and a second conditional $\pi/2$ rotation gate (see \ref{ax:ptransfer}).
The timescale of the unconditional $Z$ gate is the inverse of the Larmor frequency of the nuclear spin, which is significantly faster than the conditional rotation.
For one of the conditions in our experiment, the conditional $\pi/2$ rotation gate can be implemented with approximately $T_\mathrm{CR}=\SI{4.2}{\micro\second}$.
Additionally, four $\pi/2$ pulses on the electron spin are required at the front and end of the sequence and between the conditional and unconditional gates.
Thus, we estimate that a population transfer can be implemented with a gate time of \SI{8.7}{\micro\second}.
Note that this gate time depends on the strength of the off-axis field and the resultant effective coupling; thus, a faster gate is available.
We chose this moderate operation speed mainly owing to the timing resolution, as previously indicated.

Nuclear-assisted protocols require the nuclear spin to remain alive during electron spin manipulations, including initialization.
The nitrogen spin state is disturbed by the laser illumination that is used to initialize the electron spin~\cite{Fuchs2008, Neumann2009, Fuchs2010, Fuchs2012}; the disturbance is smaller with a smaller off-axis magnetic field~\cite{Neumann2010QND}.
Thus, an off-axis field may deteriorate the sensitivity of the nitrogen-assisted repetitive readout.
Our method requires a small off-axis field; a further study is required to determine a magnetic field condition that is sufficiently strong for nuclear spin manipulations, yet weak enough to preserve the nuclear spin.
Alternatively, applying the off-axis field only during the population transfer and switching it off during laser illumination can be another solution.

Note that an off-axis magnetic field degrades the photoluminescence intensity of NV centers~\cite{Tetienne2012NJP} and thus affects the sensitivity.
This effect is mitigated in fields far from LAC but depends on the off-axis field strength.
A careful design of the magnetic field is necessary to obtain a better sensitivity.

\section{Conclusions} \label{sec:conclution}

The nitrogen nuclear spins of NV centers are resources for quantum sensing; however, manipulating nuclear spins requires an long gate time.
We demonstrated the generation of effective coupling by an off-axis field and controlling the nitrogen spin via dynamical decoupling.
The estimated gate time based on our method is \SI{8.7}{\micro\second} for a population transfer, which is significantly faster than the direct operations by resonant pulses on the nitrogen spins.
Moreover, our method is available for an ensemble of NV centers and not only for individual NV centers, which enables nuclear-assisted protocols regarding an ensemble of NV centers.

\ack
\addcontentsline{toc}{section}{Acknowledgments}
This work is supported by the MEXT Quantum Leap Flagship Program (MEXT Q-LEAP), Grant Numbers JPMXS0118067395.
KA received funding from JST PRESTO (Grant Number JPMJPR20B1).
HI received funding from JST PRESTO (Grant Number JPMJPR17G1).

\subsection*{Author Contributions}
KM, KA, and MH conceived the project.
The experiments were conducted using KM and HT.
Data analyses and numerical simulations were conducted using KM, HT, and IF.
The diamond samples were prepared using HI and SO.
This manuscript was prepared by KM, IF, and KA, with review contributions from all the other authors.
Overall supervision was performed using KA, TI, and MH.

\subsection*{Data Availability}
The data supporting the findings of this study are available from the corresponding authors upon request.

\subsection*{Code Availability}
The codes used in this study are available from the corresponding authors upon request.

\appendix

\section{Population transfer from the electron to nuclear spin} \label{ax:ptransfer}

\begin{figure}[htbp]
  \centering
  \includegraphics[width=.45\textwidth]{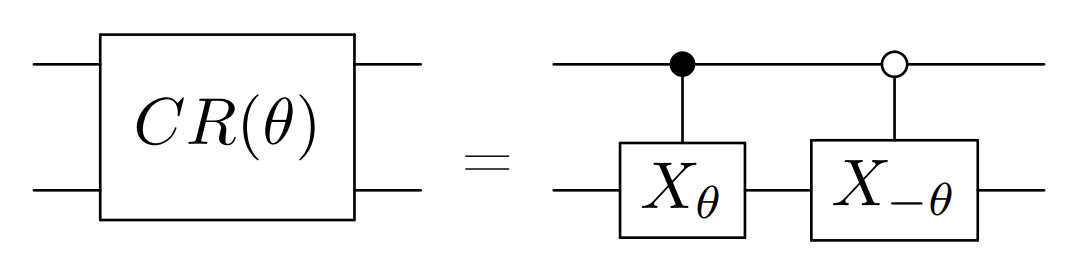}
  \caption{Conditional rotation gate with a rotation angle $\theta$.}
  \label{figax:cr}
\end{figure}

\begin{figure}[htbp]
  \centering
  \includegraphics[width=.45\textwidth]{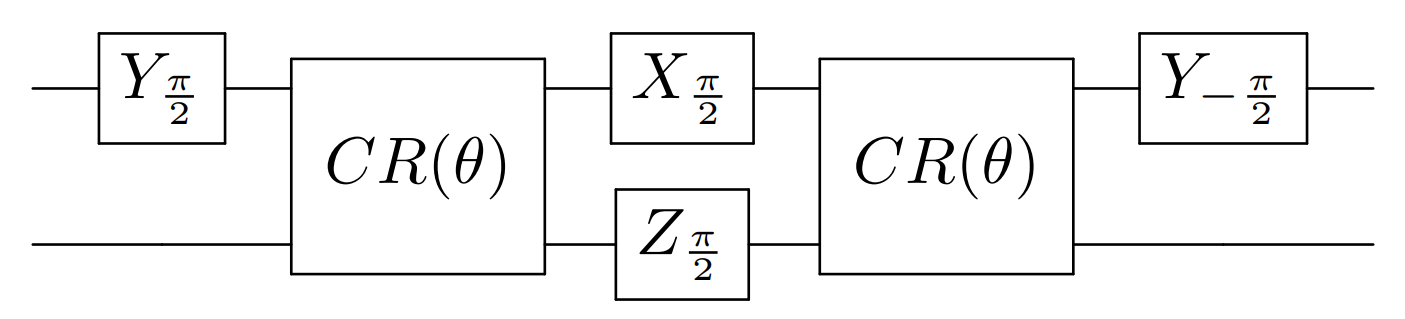}
  \caption{Population transfer from the electron spin to nuclear spin, based on the conditional rotation gate. When the conditional rotation angle $\theta = \pi / 2$, population transfer is implemented.}
  \label{figax:transfer}
\end{figure}

The population transfer we discussed in the main text is a variant of a nuclear spin operation used in Taminiau~\textit{et~al.}~\cite{Taminiau2014}.
This operation comprises three $\pi/2$ gates on the electron, two conditional rotation gates, and a $Z/2$ gate on the nuclear spin (Fig.~\ref{figax:cr} and \ref{figax:transfer}).
Note that (unconditional) $Z$ gates on the nuclear spin can also be implemented by dynamical decoupling~\cite{Taminiau2014}.
Let $\theta$ be the rotation angle of the conditional rotation gates.
The propagator is the following:
\begin{equation}
  \hat U_\mathrm{trans} = \left(\matrix{
    1 & 0 & 0 & 0 \cr
    0 & -i \cos\theta & \sin\theta & 0 \cr
    0 & \sin\theta & -i \cos\theta & 0 \cr
    0 & 0 & 0 & -1 \cr
    } \right).
\end{equation}
Assuming the quantum state before the transfer is $|\psi\rangle = \left(c_0 |0_e\rangle + c_1 |1_e\rangle\right)\otimes|0_n\rangle$, this operation converts it into $\hat U_\mathrm{trans}|\psi\rangle = \left( c_0|0_e\rangle - i c_1 \cos\theta |1_e\rangle \right) \otimes |0_n\rangle + c_1 \sin\theta |0_e 1_n\rangle$.
By tracing out the electron spin state, the nuclear spin state is $\tilde{p}_0 |0_n\rangle\langle0_n| + \tilde{p}_1 |1_n\rangle\langle1_n|$, where $\tilde{p}_0 = |c_0|^2 + |c_1|^2 \cos^2\theta$ and $\tilde{p}_1=|c_1|^2\sin^2\theta$.
Thus the electron spin population is transferred to the nuclear spin when $\theta=\pi/2$.


\section*{References}
\addcontentsline{toc}{section}{References}
\bibliography{references}
\bibliographystyle{iopart-num}

\end{document}